\documentclass[letter]{aa} 
\usepackage{graphicx}
\usepackage{txfonts}
\usepackage{lipsum}
\usepackage{subcaption}        
\usepackage{comment}
\usepackage{wrapfig}
\usepackage{lscape}            
\usepackage{placeins}         
\usepackage[colorlinks=true,linkcolor=blue,citecolor=blue,filecolor=blue]{hyperref}
\usepackage[normalem]{ulem}
\newcommand{\linkorcid}[1]{\href{https://orcid.org/#1}{\includegraphics[width=8pt]{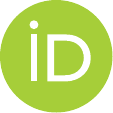}}}
\usepackage{placeins}

\begin{document}

   \title{Precursors in tidal disruption events: repeating, fast, and AGN-hosted TDEs}
   \titlerunning{Precursors in TDEs}

     \author{Patrik Mil\'an Veres \inst{\ref{rub}}\linkorcid{0000-0002-9553-2987}}        

   \institute{Ruhr University Bochum, Faculty of Physics and Astronomy, Astronomical Institute (AIRUB), Universitätsstraße 150, 44801 Bochum, Germany
              \email{veres@astro.ruhr-uni-bochum.de} \label{rub}       
        }
   \date{}

 
  \abstract
   {Tidal disruption events (TDEs) are rare transients that provide important insights into the physics of galactic nuclei. A recently identified feature in their optical light curves is the presence of early bump-like structures (‘precursors’) that appear before the onset of the main flare or during its rise.}
   {We aim to build and study the first sample of precursor TDEs in order to improve our understanding of these features, which could be key to revealing the origin of the optical emission in TDEs.} 
   {We compiled all known precursor TDEs from the literature, searched for additional candidates, and analyzed them as a sample.}
   {We find that precursor TDEs predominantly fall within the repeating TDE, fast TDE, and TDE in active galactic nucleus (AGN) subclasses. We reveal a positive correlation between the occurrence time of the precursors relative to the main peak and the central black hole mass.}
   {We suggest that the precursors appear due to interactions between the incoming stellar debris and the disk or leftover material from an earlier disruption (repeating and fast TDEs) or a stable pre-existing disk (TDEs in AGNs). Precursors are therefore potentially key signatures of repeating partial TDEs in previously quiescent galaxies.}

   \keywords{Galaxies: nuclei --
                Galaxies: active -- Accretion, accretion disks -- Black hole physics
               }

   \maketitle
   \nolinenumbers

\section{Introduction}
Tidal disruption events (TDEs) are rare transients that occur when a star passes too close to a supermassive black hole (SMBH), whose tidal forces tear the star apart \citep{1988Natur.333..523R}. About half of the star’s material is accreted onto the black hole, generating a luminous outburst which can range from radio to X-ray wavelengths. TDEs are primarily identified through optical surveys \citep[e.g., the Zwicky Transient Facility, ZTF;][]{2019PASP..131a8002B} and X-ray instruments \citep[e.g., eROSITA;][]{2021A&A...647A...1P}. The vast majority of these events exhibit one optical flare. However, this contrasts with theoretical predictions that partial disruptions, in which a bound stellar core survives and may produce recurrent flares, should occur frequently \citep{2023MNRAS.524.3026B}. Repeating partial tidal disruption events (pTDEs) can be envisioned as a natural outcome of the Hills capture mechanism \citep{1988Natur.331..687H}, in which a binary star is disrupted by a SMBH, capturing one component on a highly eccentric orbit susceptible to repeated partial disruptions. At each pericenter passage, the star loses a fraction of its mass to the SMBH, producing episodic stripping and recurrent flares on month-to-year timescales. To date, only a handful of pTDE candidates have been discovered: AT2020vdq \citep{2020vdq}, ASASSN$-$14ko \citep{2021ApJ...910..125P,14ko2,14ko3,2023ApJ...956L..46H,14ko_dance}, eRASSt J045650.3$-$203750 \citep{2023A&A...669A..75L,2024A&A...683L..13L}, RX J133157.6-324319.7 \citep{2022RAA....22e5004H},  AT2018fyk \citep{fyk1,fyk2,2024ApJ...970..116W,2024ApJ...971L..31P}, AT2019avd \citep{avd1,2021ApJ...920...56F,avd2,avd3,avd4}, AT2019azh \citep{azh_Hinkle}, AT2022dbl \citep{dbl_unlucky,dbl,dbl_imp}, AT2021aeuk \citep{at2021aeuk_jingbo}, AT2022agi (in ULIRG F01004-2237) \citep{2017NatAs...1E..61T,f01_partial}, AT2019aalc \citep{veres,2025ApJ...989..173S} and AT2023uqm \citep{uqm}.

The vast majority of TDEs were discovered in previously dormant galaxies. However, TDEs are expected to occur at comparable or even higher rates in active galactic nuclei (AGN) \citep{2015MNRAS.452...69M,tdes_in_agn,2022MNRAS.514.4102M,2024MNRAS.527.8103R}. Although the number of observationally confirmed TDEs occurring in AGN (hereafter TDE-AGN) candidates remains smaller than expected, likely due to strong selection effects and the difficulty of distinguishing them from intrinsic AGN variability, as discussed by \cite{2015MNRAS.452...69M}. Identified promising candidates include PS16dtm \citep{ps16dtm,2023A&A...669A.140P}, AT2021aeuk \citep{at2021aeuk_jingbo}, AT2024kmq \citep{kmq}, AT2022agi \citep{2017NatAs...1E..61T,f01_partial} and AT2019aalc \citep{veres,2025ApJ...989..173S}. A recently identified class of spectroscopically unique flares from flaring SMBHs — referred to as Bowen Fluorescence\footnote{A line fluorescence process, that excites O\,{\sc iii} and N\,{\sc iii} lines in the optical and near-UV \citep{1934PASP...46..146B,1935ApJ....81....1B}, often detected in TDEs.} Flares \citep[BFFs,][]{Trakhtenbrot} — may represent TDEs occurring in AGNs \citep{veres,2025arXiv251210764M}. However, it is possible that the sudden change in accretion rate observed in BFFs may instead be explained by accretion disk instabilities \citep{2025ApJ...989..173S}, while the presence of BF lines could be linked to a newly launched outflow driven by this rapid increase in accretion rate \citep{Trakhtenbrot}.

The origin of the optical and UV emission in TDEs is unclear, with leading explanations including reprocessing of X-ray photons and shocks within the stellar debris streams. TDEs are characterized by a rapidly rising, nearly colorless optical flare that subsequently decays approximately as $t^{-5/3}$, following the fallback rate of the disrupted debris. According to \cite{2019ApJ...883L..17C}, the fallback rate from a pTDE is generally expected to be approximately $t^{-9/4}$ and effectively independent of the mass of the core that survives the encounter. However, three-dimensional hydrodynamic simulations predict a broader range of decline rates, spanning approximately $t^{-p}$ with $p \simeq 2-5$, depending on the stellar structure (and hence mass) and the impact parameter \citep{2020ApJ...904..100R}. TDE flares typically show a smooth rise and decline. Only a small number of TDEs have been observed to exhibit early bump-like features in their optical light curves, occurring either prior to the onset of the main flare or during its rise (for simplicity, we refer to them as 'precursors' throughout this manuscript). Possible scenarios listed by \cite{gn} in order to explain the precursor features are (i) radiation from electron recombination and cooling in unbound debris, (ii) stream-stream collision, (iii) wind-stream collision, (iv) vertical shock compression during the first passage, and (v) the shock breakout of the debris collision. Recently, \cite{14ko_dance} and \cite{2025arXiv250907535Z} came to the conclusion that the precursors detected in the TDE-AGNs ASASSN-14ko and AT2021aeuk, respectively, are the consequences of stellar stream-disk interactions. In this Letter, we study the precursors in all known cases as a sample for the first time in order to better understand their origin. Throughout the paper, we adopt a flat $\Lambda$CDM cosmological model with parameters $H_0=70\,\textrm{km\,s}^{-1}\textrm{\,Mpc}^{-1}$, $\Omega_\Lambda=0.73$, and $\Omega_\textrm{m}=0.27$ \citep{2006PASP..118.1711W}.

\section{The precursor sample}
\label{sect2}
We collected all known regular TDEs and TDE-AGNs with reported precursor features in their optical light curves from the literature. We found $13$ cases in total, including six regular TDEs, five TDE-AGNs and two periodically repeating nuclear transients likely associated with pTDEs. Additionally, after a visual inspection of the publicly available optical photometric data, we identify the regular TDEs AT2024pvu and AT2022dbl, as well as the TDE-AGN candidate AT2019aalc, as exhibiting comparable signatures. The precursor sample, including the TDE subclass, the precursor occurrence times and their durations, are presented in Appendix \ref{table:1}. We requested optical differential light curves of all $16$ transients using the Forced-Photometry Services of the ZTF\footnote{\url{https://ztfweb.ipac.caltech.edu/cgi-bin/requestForcedPhotometry.cgi}}, the Asteroid Terrestrial-impact Last Alert System (ATLAS)\footnote{\url{https://fallingstar-data.com/forcedphot}} and the All-Sky Automated Survey for Supernovae (ASAS-SN)\footnote{\url{https://asas-sn.osu.edu}} and followed the recommendations outlined in the corresponding documentations of these services. We corrected for extinction by adopting the galactic extinction \citep{2011ApJ...737..103S}. The light curves around the precursor features are shown in Appendix \ref{app:comb}. We find that every precursor TDE falls into three categories: repeating TDEs, fast TDEs, and TDE-AGNs. We note that some sources fall within multiple categories: AT2019aalc, AT2021aeuk and ASASSN-14ko are candidate repeating TDE-AGNs, whilst the fast TDE AT2020zso occurred in a low-luminosity AGN and possibly AT2019qiz as well. In the following we refer to these with their \textit{conventional} association.

\subsection{Repeating TDEs}
Interestingly, five TDEs from our precursor sample exhibit multiple optical flares. In three cases, shown in Appendix \ref{app:b}, the optical light curves are characterized by two distinct flares and the precursor features were detected before the second ones. The first optical flares of AT2019azh and AT2024pvu were detected by the Catalina Real-time Transient Survey (CRTS). However, because of the survey’s low cadence or the absence of observations closely preceding the first flares, it is not possible to determine whether a precursor event occurred before them. The repeating nature of AT2019azh was studied by \cite{azh_Hinkle} who revealed the first flare was detected in $2005$. The precursor of the second optical flare was reported by \cite{lli} but the feature was also noted by \cite{azh_Sjoert} and \cite{azh} who referred to it as a $\approx15$\,day long slowly rising or plateau phase of the optical light curve. The precursor appears in the bolometric light curve as well \citep[see Appendix B in][]{Hammerstein23}. Its optical emission declines very fast, similar to AT2020wey discussed later. AT2022dbl is the most well-studied pTDE to date \citep[see:][]{dbl_unlucky,dbl,dbl_imp}. The precursor feature is visible only prior the second flare. The discovery and classification of AT2024pvu were reported in \cite{2024TNSAN.221....1S}. The optical light curve of it was published by \cite{Nuria} who studied the polarization features of the TDE, which properties are comparable to that of other regular TDEs. We found that a flaring episode of AT2024pvu was detected in $2006$. ASASSN-14ko and AT2023uqm have numerous optical flares. Early bumps and re-brightenings were detected prior and after the flares of ASASSN-14ko \citep{14ko_dance}. \cite{uqm} find evidence for enhanced pre-flare emission prior to the first flare of AT2023uqm. Similar precursor features can be seen prior to later flares as well, most clearly before the second flare that we plot in Appendix \ref{app:comb}.

\subsection{Fast TDEs}
Five TDEs that were detected with only a single flare also exhibit an early bump in their optical light curves. Interestingly, these are primarily known as fast TDEs i.e., a subclass of TDEs exhibiting fast-declining optical emission following the classification by \cite{wey}. AT2020wey is the first discovered fast TDE \citep{wey}. The decline rate within the first $20$\,days after peak was faster than for any previously measured TDEs. The power-law fit for the decay of the bolometric light curve resulted in a power-law index of $p=-2.84 \pm 0.18$. AT2019qiz shares several properties with AT2020wey \citep{wey}. Its bolometric light curve decays with $p=-2.51 \pm 0.03$ \citep{wey}. AT2023lli was classified as a fast TDE by \cite{lli} and its bolometric light curve decays with $p\approx-4.1$ making it the fastest-declining TDE known to date \citep{lli} together with AT2019mha. Namely, AT2019mha's decline was similarly fast with $p\approx-4$ \citep{vanVelzen20}. The light curve of AT2020zso shows a break during the rising phase $12$\,days before the peak, whereafter the slope of the rise significantly flattened \citep{zso}.

\subsection{TDE-AGNs}
We investigated the known TDE-AGN candidates and found that precursors prior to their primary optical flares were commonly detected. In the following, we describe the  temporal behavior of the TDE-AGNs in our sample. AT2021aeuk was classified as a pTDE in a Seyfert-1 galaxy by \cite{at2021aeuk_jingbo}. The precursor peaked at $L \sim 4 \times 10^{43}$\,erg/s (in ATLAS c-band) which is the highest among the entire precursor sample. Due to the enhanced optical variability between the two flares, we cannot explicitly conclude if a precursor was presented before the second flare, too. AT2019aalc is hosted by a Seyfert-1 galaxy as well and exhibited two distinct luminous optical flares. \cite{veres} suggested a repeating pTDE-AGN scenario to explain the optical re-brightening and the spectroscopic features such as the variable Bowen fluorescence and high-ionization coronal lines detected in this transient. Due to the quasi-periodic oscillations (QPOs) detected between the two main flares, we cannot conclude if a precursor appeared before the second flare, similar to AT2021aeuk. AT2024kmq was classified as a TDE-AGN by \cite{kmq} who also discusses its precursor. AT2019ahk occurred in a Seyfert galaxy showing weak AGN activity prior to the transient \citep{19bt}. Its precursor was first noted by \cite{gn}, but we note that it is only visible in the bolometric light curve published by \cite{19bt}. Therefore, we do not include this TDE in Appendix \ref{app:comb}. AT2022fpx was identified as a TDE and its host is not unambiguously classified. While its infrared colors likely indicate the presence of an AGN, SED fitting implies no significant AGN contribution to the host light \citep{fpx_1}. The low-level but clearly detected infrared variability prior to the transient \citep{fpx_devil}, however, confirms the presence of a (low-luminosity) AGN. AT2018gn is known as a TDE with precursor that occurred in a dusty environment, likely in an AGN \citep{gn}.

\section{Discussion and Conclusions}
The nature of the precursors is yet unclear. \cite{14ko_dance} stated that the early bumps and re-brightenings detected in the TDE-AGN ASASSN-14ko could potentially be explained by the energy dissipation from the plunging of the tidally stripped stream through an expanded accretion disk. In agreement with this scenario, the precursor of the TDE-AGN AT2021aeuk was explained with debris-disk interactions using hydrodynamical simulations by \cite{2025arXiv250907535Z}. We found that other TDE-AGNs also exhibit this feature. Notably, the QPOs detected in AT2019aalc were interpreted as the Lense–Thirring precession \citep{1918PhyZ...19..156L} of a misaligned disk hit by the stellar debris in an inclined orbit \citep{Nuria}. This provides an important evidence for the presence of stellar debris-disk interactions in this system. Because the innermost stable circular orbit (ISCO) scales linearly with BH mass ($R_{\mathrm{ISCO}} \propto M_{\mathrm{BH}}$, \cite{1972ApJ...178..347B}), the radial extent of the inner accretion flow increases with $M_{\mathrm{BH}}$. As a result, the returning debris intersects the disk at larger radii and therefore at earlier \textit{relative} times, potentially producing earlier or more pronounced precursor emission. We collected the black hole masses for our precursor sample from literature (listed in Appendix \ref{table:1}). When available, we adopted black hole masses derived from reverberation mapping \citep{1982ApJ...255..419B,2004ApJ...613..682P}; otherwise, we used estimates based on stellar velocity dispersion \citep{2020ARA&A..58..257G}, and, when neither measurement was provided (i.e., not possible to derive from existing data), we relied on galaxy-mass scaling relations \citep{2020ARA&A..58..257G}. For consistency, we recomputed the black hole masses for three instances (AT2022dbl, AT2023lli and AT2024kmq), those which previously relied on earlier scaling relations, using the relations of \cite{2020ARA&A..58..257G} by adopting the velocity dispersion and total stellar mass estimates from the literature \citep{dbl_unlucky,lli,gn,kmq}. For AT2024pvu, we estimated the black hole mass using galaxy-mass scaling relation, for which we took the total stellar mass from the Blast\footnote{\url{https://blast.scimma.org}} transient's host galaxy server. A log--log fit for the occurrence time versus the central black hole mass using orthogonal distance regression yields
\[\log\left(\frac{M_{\rm BH}}{M_\odot}\right) = (1.03 \pm 0.26)\,\log\left(\frac{t_{\rm occ}}{{\rm days}}\right) + 5.04,\] The correlation between black hole mass and precursor occurrence time is moderate to strong: the Pearson correlation in log-space is $0.687$ ($p = 0.006655$), while the Spearman rank correlation is $0.833$ ($p = 0.000217$), indicating a statistically significant positive trend at the level of $3.7$\,$\sigma$.
   \begin{figure}[h!]
   \centering
   \includegraphics[width=\hsize]{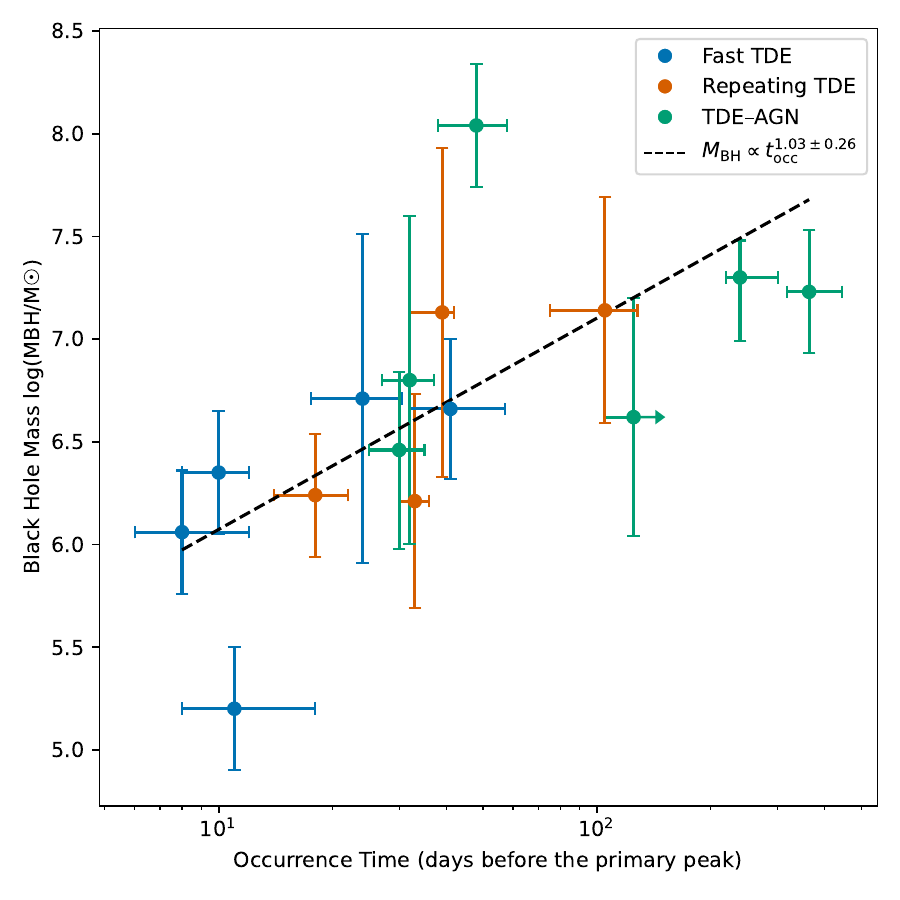}
      \caption{A positive trend between the earlier occurrence time of the precursors and the mass of the central SMBH. We define the occurrence time when the optical emission reached its highest value during the precursor flare and the x-axis uncertainties represent the duration of the precursor. The black hole mass estimates are explained in the body text in detail. We excluded AT2022fpx from the fit due to lack of observations around the precursor.}
         \label{fig:mass}
   \end{figure}
   
In fact, every\footnote{Unfortunately, only sparse sampling is available for AT2020vdq during the rise of its second flare while other reported pTDEs in literature exhibited X-ray-only re-brightenings} repeating TDEs that we could investigate exhibits a precursor. While AT2022dbl is a spectroscopically confirmed repeating pTDE, the cases of AT2019azh and AT2024pvu differ since no spectroscopic observations were performed during their first flares. This makes a pTDE scenario questionable, moreover, such long periods ($>10$\,years) require an extremely eccentric and therefore unstable orbit. One possible solution is the Hills mechanism which can result in highly eccentric but stable orbits suitable for repeated partial disruptions. \citet{f01_partial} successfully accounted for the decade-separated recurring TDEs in ULIRG F01004–2237 through the Hills mechanism. Regardless of whether AT2019azh and AT2024pvu are truly pTDEs akin to AT2022dbl, or whether their double flares result from two individual TDEs, the precursors may be linked to the repeating-flare behavior of these events. We suggest that when the returning stellar stream strikes a pre-existing disk (or leftover debris), shocks dissipate a fraction of its kinetic energy, part of which is radiated to produce the observed precursor feature. This scenario is furthermore supported by the early radio detection of AT2019azh approximately $10$\,days before the optical peak, which is the only radio detection of a thermal TDE at such early times \citep{azh_radio}. The early radio detection may be attributed to shocks produced by the interaction between the stellar debris and the fossil disk.

The scenario discussed above may explain why repeating TDEs were detected with precursors, while single TDEs typically do not display precursor features. We found only five single TDEs with precursors out of every known TDE. We note that these events rank among the most rapidly fading TDEs identified to date. In particular, AT2023lli and AT2019mha exhibit the fastest declines in their bolometric light curves, followed by AT2020wey and AT2019qiz. While AT2020zso is among the fastest TDEs in the sample of \citet{Hammerstein23}, apart from AT2020riz, only the four aforementioned events exhibit comparable e-folding decay times. The five single-flare TDEs in our precursor sample exhibit decline rates significantly steeper than the canonical $t^{-5/3}$ behavior but in agreement with the expectations of \citet{2020ApJ...904..100R} for pTDEs i.e., a decay of the form $t^{-p}$ with $p \simeq 2-5$. These events may not represent full disruptions or even the initial flares, as typically assumed for newly discovered cases, but rather secondary flares resulting from partial disruptions. The stellar debris-(fossil) disk interaction scenario provides an explanation for the precursors detected in the fast TDE cases as well. This is in agreement with the similar properties observed between the precursors of repeating TDEs and those of fast TDEs, both in terms of their timing relative to the main peak and their duration.

To summarize, we suggest that TDE precursors appear due to interactions between the incoming stellar debris and the disk or leftover material from an earlier disruption (repeating TDEs) or a stable pre-existing disk (TDE-AGNs). Fast TDEs whose decline rates match those predicted in pTDE simulations may not represent full disruptions or not even first flares, and their precursors can be interpreted in the same framework as in repeating TDEs. Importantly, based on the theoretically expected pTDE rates, \cite{2023MNRAS.524.3026B} proposed that many of the regular TDEs classified as total disruptions in the standard framework are in fact pTDEs. This is in agreement with our proposed explanation for the nature of fast TDEs. Early triggering of follow-up observations for newly discovered TDEs, including in the radio band will be crucial to constrain the origin of precursors. Especially, the third flare of AT2022dbl is expected to occur in early $2026$ and will be detectable according to the light curve simulations of \cite{dbl_3fl} .Other repeating TDEs in our sample are likewise expected to undergo future flaring episodes (see Table 2 in \cite{2025arXiv251119016Q}).

\begin{acknowledgements}
I thank the anonymous referee for their constructive suggestions on our Letter. I thank for Anna Franckowiak (RUB) for fruitful discussion and for commenting the manuscript. PMV acknowledges the support from the DFG via the Collaborative Research Center SFB1491 \textit{Cosmic Interacting Matters - From Source to Signal}. I acknowledge using a color-blind friendly palette \citep{2008SPIE.6807E..0OI} in the plots.
\end{acknowledgements}

\bibliographystyle{aa}
\bibliography{biblio}
    
\onecolumn
\appendix

\section{The precursor sample}
\begin{table}[h!]
\caption{The precursor sample. $t_{\mathrm{occ.time}}$ denotes the occurrence time of the precursor relative to the primary peak, while the associated uncertainties represent the precursor duration. The parameter $p$ gives the power-law index of the bolometric light curve decay for fast TDEs.}      
\label{table:1}   
\centering      
\resizebox{\columnwidth}{!}{%
\begin{tabular}{c c c c c c c c}     
\hline\hline               
ID & TDE type & $t_{\mathrm{occ.time}}$ (days) & precursor ref. & $\log(M_{\rm BH}/M_\odot)$ & M$_{\rm BH}$ ref. & $p$ & $p$ ref. \\
\hline
AT2019azh & repeating & $18 \pm 4$ & \citet{lli}, \citet{2024ApJ...969..104F} & $6.24 \pm 0.30$ & \citet{Mummery23} & & \\
AT2024pvu & repeating & $39^{+3}_{-7}$ & this work & $7.13 \pm 0.80$ & this work & &  \\
AT2022dbl & repeating & $18 \pm 3$ & this work & $6.21 \pm 0.52$ & this work & &  \\
AT2023uqm & repeating & $105^{+23}_{-30}$ & this work, \citet{uqm} & $7.14 \pm 0.55$ & \citet{uqm} &  &  \\
AT2020wey & fast & $11 \pm 3$ & \citet{wey} & $5.20 \pm 0.52$ & \citet{Mummery23} & $-2.84 \pm 0.18$ & \citet{wey} \\
AT2023lli & fast & $39^{+16}_{-9}$ & \citet{lli} & $6.66 \pm 0.34$ & this work & $-4.1$ & \citet{lli} \\
AT2019qiz & fast & $10 \pm 2$ & \citet{gn} & $6.35 \pm 0.30$ & \citet{Mummery23} & $-2.51 \pm 0.03$ & \citet{wey} \\
AT2019mha & fast & $24 \pm 7$ & \citet{gn} & $6.71 \pm 0.80$ & \citet{Mummery23} & $-4.0^{+0.7}_{-0.5}$ & \citet{vanVelzen20} \\
AT2020zso & fast? TDE-AGN & $8^{+4}_{-2}$ & \citet{gn} & $6.06 \pm 0.30$ & \citet{Mummery23} & $-2.12^{+0.28}_{-0.41}$ & \citet{Hammerstein23} \\
AT2022fpx & TDE-AGN & $>125$ & \citet{fpx_1} & $6.62 \pm 0.58$ & \citet{fpx_1} &  &  \\
AT2018gn & TDE-AGN & $30 \pm 5$ & \citet{gn} & $6.46^{+0.38}_{-0.48}$ & \citet{gn} &  &  \\
AT2024kmq & TDE-AGN & $48 \pm 10$ & \citet{kmq} & $8.04 \pm 0.30$ & this work &  &  \\
AT2019ahk$^{a}$ & TDE-AGN & $32 \pm 5$ & \citet{gn} & $6.80 \pm 0.80$ & \citet{19bt} &  &  \\
AT2019aalc & repeating TDE-AGN & $364^{+81}_{-45}$ & this work & $7.23 \pm 0.30$ & \citet{Lancel} &  &  \\
AT2021aeuk & repeating TDE-AGN & $239^{+62}_{-20}$ & \citet{at2021aeuk_jingbo} & $7.30^{+0.18}_{-0.31}$ & \citet{gleok} &  &  \\
ASASSN-14ko$^{b}$ & repeating TDE-AGN & $\sim 20$ & \citet{14ko_dance} & $7.86^{+0.31}_{-0.41}$ & \citet{2021ApJ...910..125P} &  &  \\
\hline                                 
\end{tabular}%
}
\footnotesize{
\textsuperscript{a} The precursor appears only in the bolometric light curve. \\
\textsuperscript{b} The precursors appear on different timescales across different flaring episodes.
}
\end{table}

\newpage
\section{Light curves of the precursor sample}

\begin{figure*}[h!]
  \centering
  \includegraphics[width=0.9\textwidth]{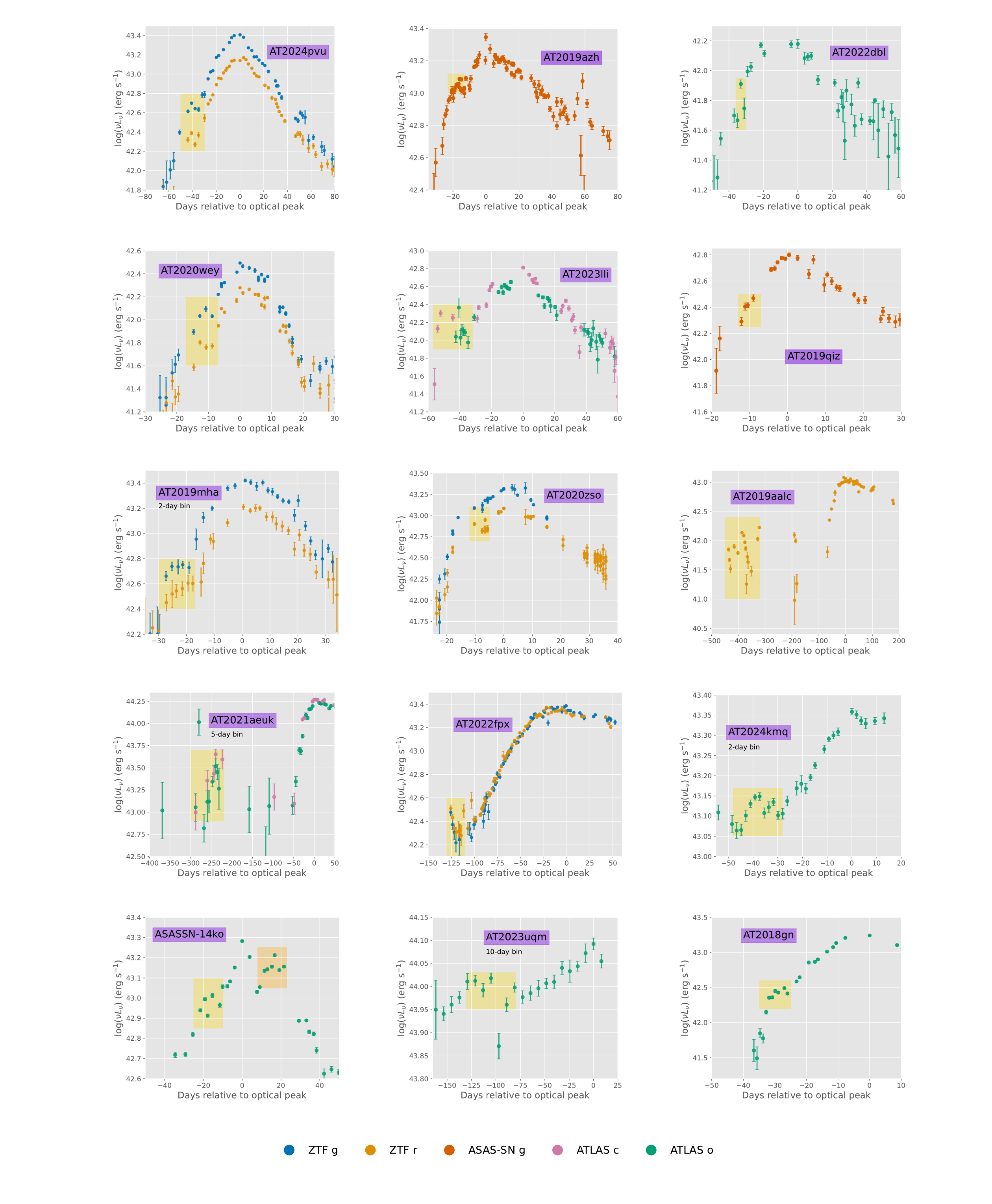}
  \caption{Optical light curves of our precursor sample. We show the differential (i.e., host-subtracted) light curves, with the only exception of AT2023uqm, whose precursor flare mainly exhibits negative fluxes (i.e., fainter than the baseline emission); in this case, we show the aperture photometry.  The precursor features are highlighted with gold areas. For ASASSN-14ko — in which case the re-brightening during decline was studied earlier in detail — we also highlight this phase with an orange shaded area. Days are always in rest-frame. The light curves were binned only for intranight observations to improve the signal-to-noise ratio, unless stated otherwise in the figure annotations.}
  \label{app:comb}
\end{figure*}

\FloatBarrier
\section{Long-term light curves of the repeating TDEs}
\label{app:b}
We present the long-term optical light curves of the repeating TDEs with double flares. WISE infrared light curves are also shown on top, retrieved using \textsc{timewise} \footnote{\url{https://timewise.readthedocs.io/en/latest/}}.
\begin{figure}[h!]
  \centering
  \includegraphics[width=0.7\textwidth]{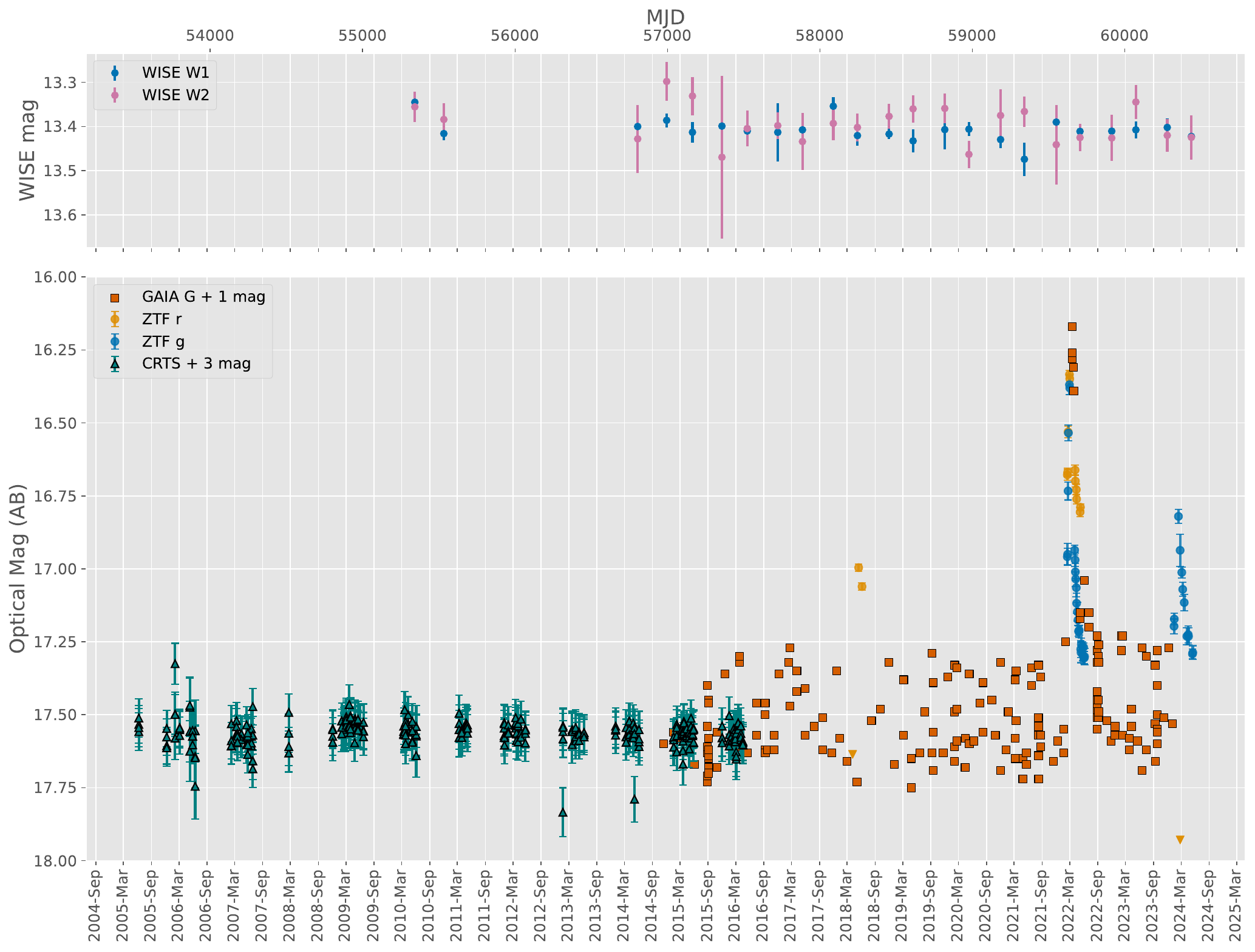}
  \caption{Long-term optical light curve of the spectroscopically confirmed repeating pTDE AT2022dbl. We plotted ZTF aperture magnitudes in order to be consistent with CRTS and Gaia photometries which are not corrected for baseline. No enhanced infrared emission was detected by WISE after the first optical flare while no data is available after the second.}
  \label{app:dbl}
\end{figure}

\begin{figure}[h!]
  \centering
  \includegraphics[width=0.7\textwidth]{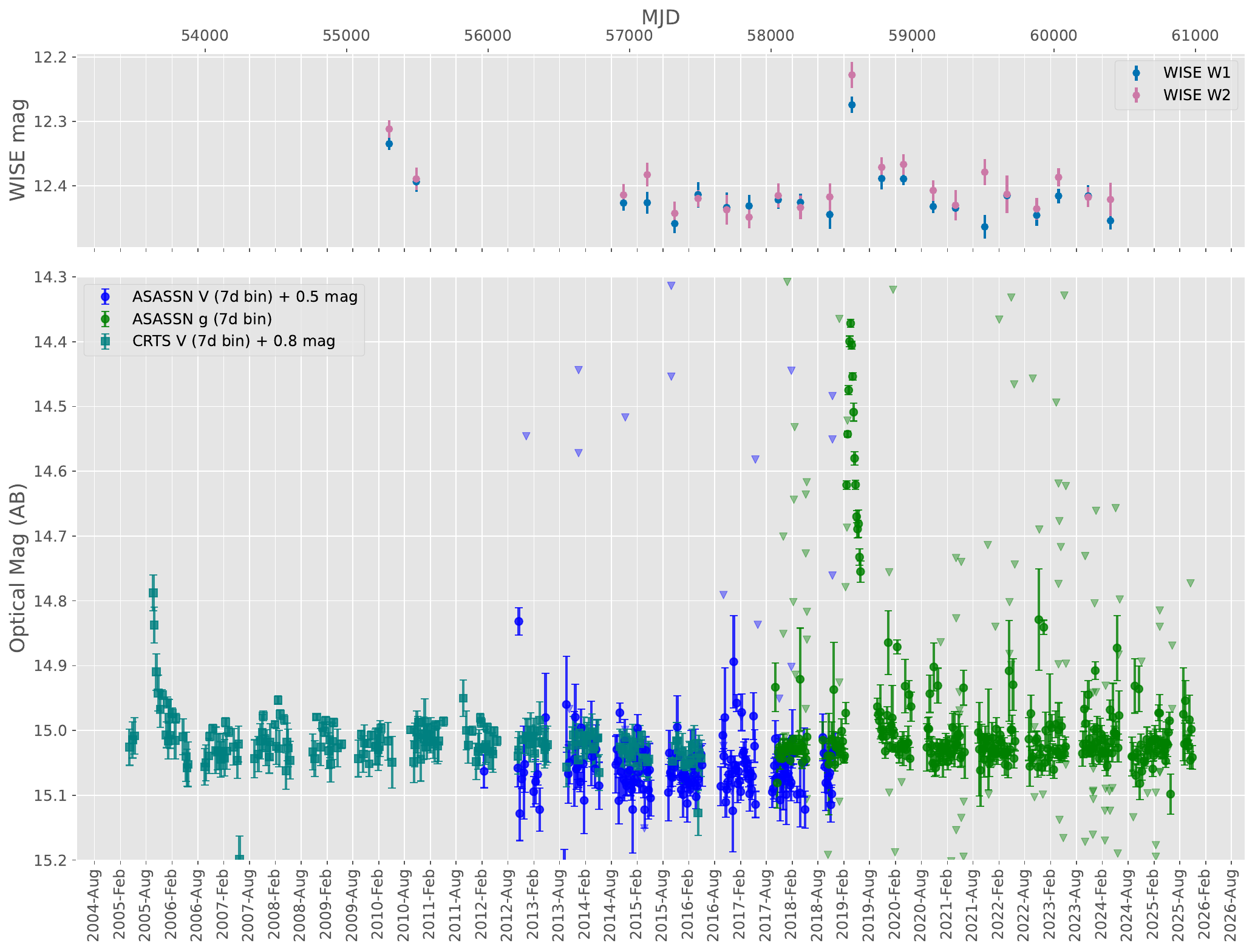}
  \caption{Long-term optical light curve of the repeating TDE AT2019azh presented in aperture magnitudes. The second flare appeared $14$ years after the first one detected by CRTS. Two individual TDE scenarios and a repeating pTDE scenario are discussed in more detail in the main text. The second flare was accompanied by an IR dust echo flare whilst no data is available after the first flare for comparison. }
  \label{app:azh}
\end{figure}

\begin{figure}[h!]
  \centering 
  \includegraphics[width=0.7\textwidth]{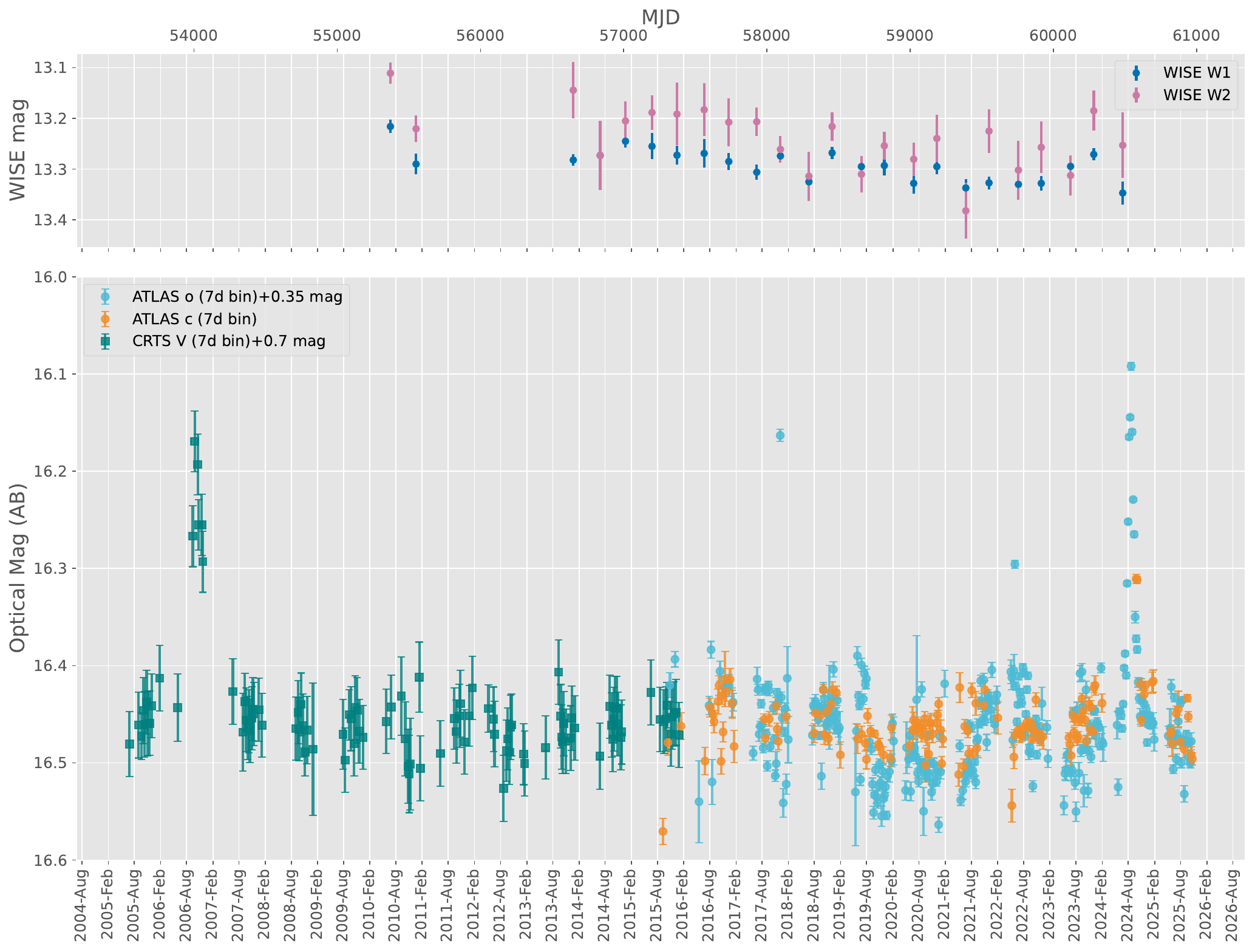}
  \caption{Long-term optical light curve of the repeating TDE AT2024pvu presented in aperture magnitudes. The second flare appeared $18$ years after the first one detected by CRTS. Two individual TDE scenarios and a repeating pTDE scenario are discussed in more detail in the main text. Unfortunately, no WISE observations were obtained after the second optical flare due to the satellite’s shutdown.}
  \label{app:pvu}
\end{figure}

\end{document}